\documentclass[conference]{IEEEtran}
\usepackage{cite}
\usepackage{amsmath,amssymb,amsfonts}
\usepackage{graphicx}
\usepackage{textcomp}
\usepackage{xcolor}
\def\BibTeX{{\rm B\kern-.05em{\sc i\kern-.025em b}\kern-.08em
    T\kern-.1667em\lower.7ex\hbox{E}\kern-.125emX}}

\usepackage{algorithm}
\usepackage[noend]{algpseudocode}
\def\BState{\State\hskip-\ALG@thistlm}

\begin{document}

\title{Bluetooth Low Energy mesh network for power-limited, robust and reliable IoT services}

\author{\IEEEauthorblockN{Davide Villa}
\IEEEauthorblockA{\textit{Northeastern University} \\
Boston, MA, USA \\
\{villa.d\}@northeastern.edu}
\and
\IEEEauthorblockN{Chih-Kuang Lin}
\IEEEauthorblockA{\textit{Collins Aerospace} \\
Cork, Ireland \\
\{chih-kuang.lin\}@collins.com}
\and
\IEEEauthorblockN{Adam Kuenzi, Michael Lang}
\IEEEauthorblockA{\textit{Carrier Corporation} \\
Salem, Oregon, USA \\
\{adam.kuenzi, michael.lang\}@carrier.com}
}

\maketitle

\thispagestyle{plain}
\pagestyle{plain}

\begin{abstract}
Bluetooth Low Energy (BLE) is an emerging wireless technology created for short-range control and monitoring applications that is becoming increasingly widespread among the Internet of Things (IoT) services because of its low-cost and\linebreak low-energy consumption. In this paper, we propose a novel neighbor discovery scheme and failure recovery techniques for multi-path and single-path low-power and reliable BLE networks. By exploiting energy-efficient access control and fast and robust routing ideas with adaptive failure recovery, the proposed methods outperform the well-known flooding approach used by the BLE Mesh standard. We show varying improvements in packet latency and power consumption in event-driven simulations as network topology and traffic changes. The failure recovery approaches proposed are optimized and demonstrated during the simulations, showing the varying of the overall failure recovery latency and node power consumption in different use cases.
\end{abstract}
\vspace*{2mm}

\section{Introduction}
Internet of Things (IoT) applications often face two challenges: long battery lifetime and high reliability \cite{ckbib1} \cite{ckbib2}. That leads to the use of Bluetooth Low Energy (BLE) technology for IoT services, e.g. access control, object tracking, monitoring \cite{ckbib3}. It is popular because of the prevalence of BLE devices, interoperability with mobile phones, high throughput, low power, and low cost. Mesh networks enable M-to-M communications among local nodes. This technology builds a large wireless network via multi-hop delivery and it is therefore reasonable to consider the BLE mesh for IoT applications.\\
In 2017, Bluetooth Mesh networking standard was published \cite{ckbib4}. It assumes the use of non-power limited nodes, named Friend nodes, to build the multi-hop backbone of the BLE mesh network. Other power-limited nodes connect to the backbone within single hop, building a large BLE network. Since this method adopts the friendship node concept, it is infeasible to have a BLE mesh network with all power-limited nodes. The standard also lacks an energy-efficient MAC and redundant routing, so the current BLE Mesh standard is unable to support the critical service with high reliability and long battery lifetime. This paper will investigate novel methods for a low-power, robust, and reliable BLE mesh network where most or all nodes are battery powered with no Friend nodes.\\
In the past, there have been some works to enable BLE mesh networks, e.g. flooding-based and routing-based solutions \cite{ckbib4} \cite{ckbib5} \cite{ckbib7} \cite{ckbib8}. The flooding method broadcasts packets across the network over BLE advertising channels, creating concerns, e.g. advertising channel contentions and radio resource waste. The routing method uses a routing protocol to forward data in BLE networks. Its challenges include the energy saving, and Quality of Service.
In this paper, we propose the cross layers solution meeting the challenges above. First, an energy-efficient access control policy compatible with BLE 5.0 specifications \cite{ckbib6} is proposed. That minimizes active period of a node and increases transmission reliability. In the network layer, a fast and robust multi-path routing approach is proposed with an adaptive failure recovery method. This idea uses the multi-path configuration to provide the fast recovery for a failed route, together with an adaptive recovery selection that optimizes the network performance. To evaluate these techniques, we first describe the proposed methods in Section \ref{section3}. In Section \ref{section4}, detailed simulation models and results are presented and analyzed. Finally, Section \ref{section5} discusses the conclusions.

\section{State of the art}
BLE networking solutions are classified into flooding and routing methods. The flooding approach is easy to implement with low overhead cost. In Bluetooth Mesh standard the sender uses advertising channels to broadcast data with a message cache and Time-To-Live (TTL) ensuring a relay node only to forward a message once over a limited number of intermediate hops. Receive node uses a 100\% scan mode for fast data receptions with high energy cost \cite{ckbib4}. Gogic et. al. report a flooding protocol with bounded packet delivery ratio and latency while keeping energy cost low \cite{ckbib7}. A bounded flooding solution, called BLE Mesh, is proposed in \cite{ckbib8}. It limits rebroadcasting in intermediate nodes by using local neighbors information and cost analysis of a retransmission. Routing-based approach is the other BLE networking direction. In \cite{ckbib10}, the authors propose a static tree topology for the routing over BLE. By introducing the root, intermediate, and leaf nodes and master/slave hierarchical rules, a node could send data to any node in the tree. Real Time BLE is another routing method over BLE using the master/slave subnetworks \cite{ckbib11}. It focuses on the routing performance with bounded end-to-end delay and proposes the CCCD descriptor in the GATT layer to manage concurrent connections. The authors in \cite{ckbib12} \cite{ckbib13} adopt the on-demand approach to find a route using iterative route searches.\linebreak They exploit the intermediate nodes information or a breadth-first search algorithm trading off their route efficiency for the control message overhead cost. Overall, the techniques mentioned above lack a method that provides a fast, robust, and reliable recovery for failed routes. Their redundant transmissions and resource wastes are also concerns. In the routing-based category, the problems, e.g. a single node failure, limited failure recovery features, and network performance sub-optimization, are observed. Therefore, this paper will address the abovementioned problems and make contributions as followings. 1) A novel MAC solution is introduced to optimize node energy. 2) A robust multi-path routing protocol is proposed with an adaptive failure recovery technique. These enable the fast recovery for a failed route while improving network efficiency in terms of latency and power consumption.

\section{Cross-layer protocol description}\label{section3}
In this section, we describe the main features of our methodologies while the following sections present the advantages of our design \cite{pat3} \cite{pat4}. The proposed protocol takes into account a BLE system, but the same approaches can be extended to all of wireless multi-hop networks. It includes a set of the cross-layer protocols and covers data-link and network layers. First, the data-link layer provides the link establishment through the access control and the subsequent data delivery via an advertisement-scan mechanism similar to BLE and shown in Section \ref{datalinksection}. Next, the network layer aims to create and maintain the routing paths needed to perform the data deliveries. The devised mechanisms are mainly based on the hop-count distances, i.e. the distance in terms of number of hops between nodes. After a reset, the nodes perform a neighbor discovery and exploit the Greedy Search Algorithm (GSA) to build the paths, described in Section \ref{neighbordiscovery} and \ref{GSA} respectively. Finally, when a failure is discovered, proper recovery techniques are described in \ref{failurerecoverysection} and \ref{adaptivefailuresection}.

\subsection{Access control and data session} \label{datalinksection}
As for the BLE, the physical layer consists of 40 channels with 2 $MHz$ channel spacing within the 2.4-2.4835 $GHz$ ISM radio bands. The channels are organized into two groups: 3 advertising channels (37-38-39) which provide the device discovery and link establishment; and 37 data channels for the bidirectional data streams. A node periodically advertises his availability by sending an advertisement message in each advertising channel sequentially. The period of each cycle is composed by a fixed parameter, named advertising interval ($\tau_{AI}$), plus a variable random delay ($\delta$) to randomize the advertisements deliveries increasing the chances of reception. When a packet is pending in the transmission queue, the node switches to scan mode and starts the scan interval ($\tau_{SI}$). It consists of an active mode, called scan window ($\tau_{SW}$), where the node listens for advertisements in one advertising channel, and a low-power mode, where the node waits before switching to listen in the next channel cyclically (37$\rightarrow$38$\rightarrow$39$\rightarrow$37…). When a proper advertisement is received, the node replies with a connection request message, which contains the link setup parameters, e.g. data channel and hopping sequence. Then, both ends keep the data channel open. The scan node starts sending all the data packets for that destination one by one and the receiver will reply with an acknowledgment to each one of them.
If the ack is not received, the data is retransmitted till a certain threshold beyond which the sender will switch in scan mode again to establish a new link. 
If no advertisements from interested nodes are received during a certain period of time, the sender declares the current destination failed and a failure recovery is triggered. The time needed to declare a failure might depend on several factors, e.g. interference, network age or traffic intensity. A higher value takes more time to detect a failure but might result in a less number of false positive cases. 
The advertising operations are also performed during the scan mode to avoid potential deadlocks caused by nodes that want to send packets to each other at the same time.

\subsection{Neighbor discovery}\label{neighbordiscovery}
The neighbor discovery aims at retrieving the neighboring nodes hop-count distances towards the closest head node, e.g. data sink or gateway, and their 1-hop neighbor list with correspondent information, e.g. MAC address and hop-count. These data are included in the advertisements messages and are exploited by the neighbor discovery algorithm. Its operations are described in Algorithm \ref{neighdiscalgorithm}.
At the reset, all nodes have their hop-counts undefined, i.e. $0xFF$. The sole exceptions are the head nodes which have an initial value of $0$. An advertisement is new when it is received for the first time from a neighbor not yet discovered, or when it contains different data compared to the previous one received. At lines \ref{ndaset1st} and \ref{ndaset2nd} a node may update its own hop-count ($my\_hop$) incrementing the neighbor hop-count received in the new advertisement ($adv\_hop$). The discovery timer mainly depends on the network density, i.e. the average number of 1-hop neighbors to discover, and on the channel quality for the proper advertisements reception. The higher the value, the longer the time needed by each node to complete the discovery, but also the higher the probability to correctly discover all of the neighbors. An estimation of its optimal value could be achieved running model simulations.
\begin{algorithm}[]
    \small
	\caption{Neighbor discovery}\label{neighdiscalgorithm}
	\begin{algorithmic}[1]
		\Procedure{All nodes operations}{}
		\State \textbf{switch} to scan mode to receive adv;\label{ndascan}
		\If {new adv with no $0xFF$ hop-count is received}
		\If {first adv received}
		\State \textbf{start} a discovery timer;
		\State \textbf{set} $my\_hop = adv\_hop\hspace{1pt}+\hspace{1pt}1$;\label{ndaset1st}
		\Else
		\State \textbf{reset} the discovery timer;
		\If {$my\_hop\hspace{1pt}>\hspace{1pt}(adv\_hop\hspace{1pt}+\hspace{1pt}1)$}
		\State \textbf{set} $my\_hop = adv\_hop\hspace{1pt}+\hspace{1pt}1$;\label{ndaset2nd}
		\EndIf
		\EndIf
		\State \textbf{store} the information contained in the adv, \textbf{goto} \ref{ndascan};
		\EndIf
		\If {discovery timer expires}
		\State \textbf{stop} the neighbor discovery;
		\EndIf
		\EndProcedure
	\end{algorithmic}
\end{algorithm}

\subsection{Greedy Search Algorithm}\label{GSA}
The primary objective of Greedy Search Algorithm (GSA) is to create disjoint multi-paths with the minimum number of hops where the intermediate nodes must be different among the derived multi-paths except for original source and final destination. The GSA is described in Algorithm \ref{GSAalgorithm}.
In literature, a greedy algorithm is a paradigm that iteratively derives the results step by step, potentially leading to the optimal outcome.\linebreak
\begin{algorithm}[]
    \small
	\caption{Greedy search algorithm}\label{GSAalgorithm}
	\begin{algorithmic}[1]
		\Procedure{Original source operations}{}\label{gsa.originalsourceoperations}
		\State $maxTTL=1, \enspace nPaths=0, \enspace maxPaths=K$;
		\If {$nPaths$\hspace{3pt}$<$\hspace{3pt}$maxPaths$}\label{gsa.newpath}
		\State \textbf{create} GSA msg with $TTL=maxTTL$, \textbf{goto} \ref{gsa.sendingoperations};
		\Else
		\State \textbf{end} path creation;
		\EndIf
		\If {there are neighbors to select with \textit{hop-count\hspace{2pt}$>=$\hspace{2pt}TTL} Or in \textit{TempExcluded}} \label{gsa.orignackrec}
		\State $maxTTL$\hspace{1pt}+\hspace{1pt}+$, \enspace TTL=maxTTL$, \textbf{goto} \ref{gsa.sendingoperations} \label{gsa.incresettl}
		\Else
		\State \textbf{delete} GSA message, \textbf{end} path creation;
		\EndIf
		\If {routing ack received}\label{gsa.routingackreceived}
		\State $nPaths$\hspace{1pt}+\hspace{1pt}+, \textbf{goto} \ref{gsa.newpath};
		\EndIf
		\EndProcedure
		
		\State 
		
		\Procedure{Sending operations}{}\label{gsa.sendingoperations}
		\State \textbf{select} neighbor with (min \textit{hop-count\hspace{3pt}$<$\hspace{3pt}TTL}) \enspace \& \enspace $!(\textit{PermExcluded}) \enspace$ \& $\enspace !(\textit{TempExcluded})$; \label{gsa.nexthopselection}
		\If {more than one neighbor} 
		\State \textbf{select} node best RSSI, \textbf{send} GSA msg, \textbf{goto} \ref{gsa.receivingoperations};  \label{gsa.nexthopselection2}
		\ElsIf {no neighbors available}
		\If {original source} 
		\State \textbf{goto} \ref{gsa.orignackrec};
		\Else 
		\State \textbf{send-back} a routing nack, \textbf{goto} \ref{gsa.nackrec};
		\EndIf
		\EndIf
		\EndProcedure
		
		\State 
		
		\Procedure{Receiving operations}{}\label{gsa.receivingoperations}
		\If {GSA message received \& receiver is a head}  \label{gsa.headreception}
		\If {head is in a previous path \& the original source distance is \textit{1-hop}}
		\State \textbf{send} data-link nack, \textbf{goto} \ref{gsa.llnackrec};
		\Else
		\State \textbf{send} data-link ack, \textbf{delete} GSA message, \textbf{send-back} routing ack, \textbf{goto} \ref{gsa.routingackreceived};
		\EndIf
		\ElsIf {GSA msg received \& receiver is not head}\label{gsa.otherreception}
		\If {the node is in a previous path Or it does not have any neighbors available to select}
		\State \textbf{send} data-link nack, \textbf{goto} \ref{gsa.llnackrec};
		\Else
		\State \textbf{send} data-link ack, $TTL$\hspace{1pt}-\hspace{1pt}-, \textbf{goto} \ref{gsa.sendingoperations};
		\EndIf
		\EndIf
		\If {data-link nack received} \label{gsa.llnackrec}
		\State \textbf{add} neighbor in \textit{PermExcluded} Or \textit{TempExcluded}, \textbf{goto} \ref{gsa.sendingoperations};\label{gsa.nackreceived}
		\EndIf
		\If {routing nack received} \label{gsa.nackrec}
		\State \textbf{add} neighbor in \textit{TempExcluded}, $TTL$\hspace{1pt}+\hspace{1pt}+, \textbf{goto} \ref{gsa.sendingoperations};
		\EndIf
		\EndProcedure
	\end{algorithmic}
\end{algorithm}\\
Heuristically, these approaches are not optimal, but with sufficient data they can lead to the optimal solution. The GSA uses the local nodes information of neighbor discovery and GSA itself to find the best next hops to build $nPaths$ routes.
The algorithm exploits a Time-To-Live (TTL) value stored in the GSA messages together with data-link and network acknowledgments. Additionally, each node exploits two lists: \textit{PermExcluded} which contains the neighbors to always avoid in the next hop selection; \textit{TempExcluded} which contains the neighbors to be avoided for a certain maximum TTL value.\\
At line \ref{gsa.headreception}, the head will not accept the path creation if it has a 1-hop direct route previously built with the original source. It is unlikely that the original source is not able to communicate with this head directly, but it would need to pass through a longer path. It is preferable to create more distributed paths compared to re-use the same heads to enhance the reliability. In a network with few heads, this rule might be deleted.\\
The nack sent by a non-head node at line \ref{gsa.llnackrec} contains a flag in order to inform the sender to insert its id in the \textit{PermExcluded} list, if the receiver is in a previous path or it does not have any neighbor to select, or in the \textit{TempExcluded}, if the receiver does not have any neighbor to select with that TTL value.\\
Each GSA message contains a route id number generated by the original source that uniquely identifies the current path. It will be stored in the new entry of the routing table together with the uplink and downlink next hop ids. The downstream routes are inferred from the upstream ones. The resulting routing table built by each node is shown in Table \ref{routingtablestructure}.
\begin{table}[t]	
	\caption{Routing table structure}
	\begin{center}
		\begin{tabular}{|c|c|c|}
			\hline
			\textbf{Route id} & \textbf{Next hop upstream} & \textbf{Next hop downstream} \\ \hline
			\textit{\begin{tabular}[c]{@{}c@{}}Route path\\identification\end{tabular}} & \textit{\begin{tabular}[c]{@{}c@{}}Upstream next hop id\\retrieved from GSA\end{tabular}} & \textit{\begin{tabular}[c]{@{}c@{}}Downstream next hop id\\inferred from GSA\end{tabular}} \\ \hline 
		\end{tabular}
		\label{routingtablestructure}
	\end{center}
	\vspace*{-4.5mm}
\end{table}

\subsection{Failure recovery}\label{failurerecoverysection}
The failure recovery is triggered when a sender discovers that the next hop is unreachable. Its main goals are to inform a head node about the failure and to select an alternative route.
As first failure recovery mechanism, a reconnection setup timer is used. In this window the sender keeps trying to establish a connection with the receiver. The timer value is tunable based on the environment conditions, traffic congestion and interference. After the reconnection timer has expired, the sender has two methodologies to continue the failure recovery.\\
The first technique is the multi-path approach and works for both uplink and downlink packets. When the failure is discovered by an intermediate node in the path, the current message is dropped and a notification is sent back. When the notification arrives to the original source, or if the failure has been detected directly by the original source, the current path is invalidated and the data is sent via the next disjoint path in chronological order. If there are no more paths available, the message is marked as undeliverable. When the final destination receives the packet, it checks its route id, sets the current path active and invalidates all the previous paths in chronological order.\linebreak
This updating approach relies on the disjoint multi-path policy and on the selection in chronological order of the next paths.\\
The second method is the hop-count distance based approach and works only for uplink packets. After a failure is discovered, the sender performs a neighbor discovery to retrieve up-to-date neighbor information. This operation is optional if the node relies on the previous data stored. Then, it selects the next hop based on the minimum hop-count (as GSA step \ref{gsa.nexthopselection}). The failed node id is added to the payload and then the packet is forwarded. The next hop exploits its own upstream route if possible, i.e. if the next hop is not the previous node and it is not failed. Otherwise, it starts another neighbor discovery and selection. The same process is repeated till reaching the final destination. Each intermediate node updates the entries of that route id with the new upstream and downstream next hops.

\subsection{Adaptive failure recovery}\label{adaptivefailuresection}
\begin{figure}[t]
	\centerline{\includegraphics[trim=10.8cm 6.3cm 10.8cm 6.3cm,clip,width=8.6cm]{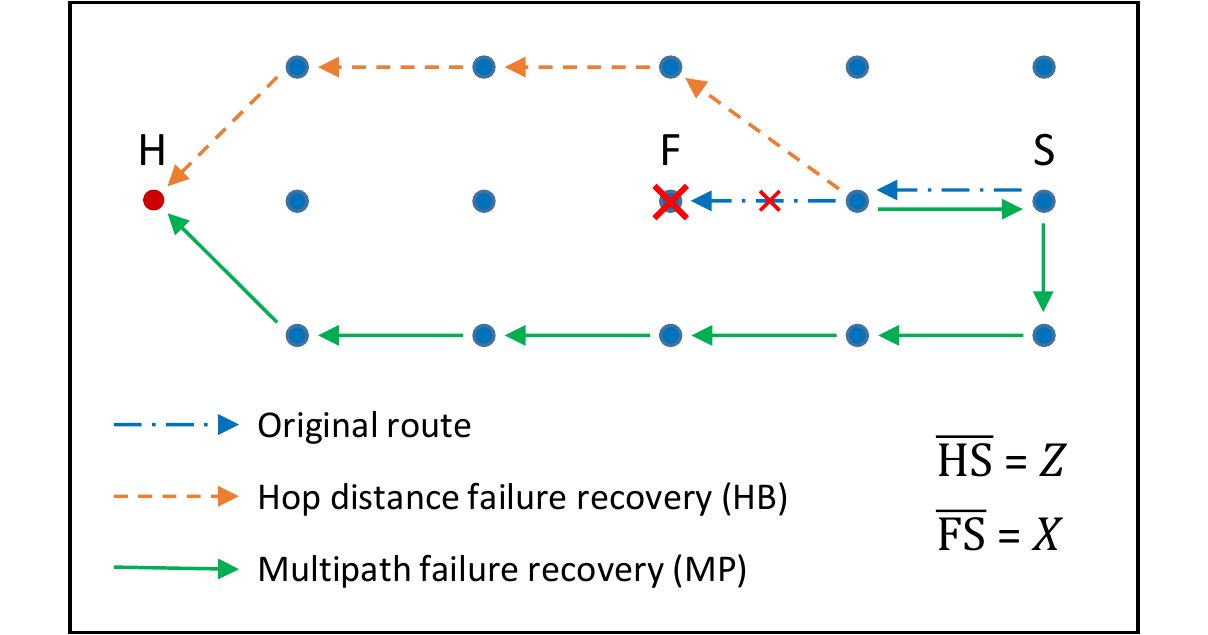}}
	\caption{Adaptive failure recovery example in a 3x5 network topology. H is the head node, F is the failed node and S is the original source.}
	\label{afrexample}
	\vspace*{-4.5mm}
\end{figure}

Combining the two failure recovery methodologies, we propose an adaptive failure recovery mechanism. It has the benefit of reducing the latency and saving resources by deriving the optimal selection in terms of recovery speed between hop-distance based (HB) and multi-path (MP) upstream failure recoveries by exploiting the node local information (Table \ref{afrsymbols}).\\
\begin{table}[b]	
	\vspace*{-4.5mm}
	\caption{Adaptive failure recovery parameters}
	\begin{center}
		\begin{tabular}{|c|c|}
			\hline
			\textbf{Notation} & \textbf{Definition} \\ \hline
			$Z$ & Hop-count distance between head and original source \\ \hline
			$X$ & Hop-count distance between failure and original source \\ \hline
			$\alpha$ & Estimated extra hops required by hop-distance recovery \\ \hline
			$\beta$ & Estimated extra hops required by multi-path recovery \\ \hline
			$r$ & Extra latency required to perform the neighbor discovery \\ \hline
			$\gamma$ & Estimated time required to perform a single-hop delivery \\ \hline
		\end{tabular}
		\label{afrsymbols}
	\end{center}
\end{table}
The HB failure recovery overall latency consists of two parts: the time to perform the neighbor discovery operations ($r$); and the time to reach the head via the node previously selected by exploiting its predefined path, equal to the distance between head and current node ($Z - X + 1$) plus an extra delay derived from the new path extra hops ($\alpha$). Similarly, the MP failure recovery latency consists of: the time to go back from the node detecting the failure to the original source ($X - 1$); and the time to reach the head node via the next alternative route, equal to the distance between original source and head ($Z$), plus the new path extra hops ($\beta$). We derived the equation (\ref{hbequation}) and (\ref{mpequation}) for HB recovery and for MP recovery respectively.
\begin{equation}\label{hbequation}
	HB = r + Z - X + 1 + \alpha
\end{equation}
\begin{equation}\label{mpequation}
	MP = Z + X - 1 + \beta
\end{equation}
The unit of measurement for both equations is the $second$. The values in hops are converted in time unit by multiplying their values by the estimated time required to perform a single hop delivery ($\gamma$). A summary of the notations together with a topology example are shown in Table \ref{afrsymbols} and in Figure \ref{afrexample}.\\
The values of $Z$ and $X$ are easily retrieved by checking the TTL of a packet, which gives information regarding the number of hops the message has crossed from the original source till the current node that has encountered the failure. The other three parameters should be estimated by the current node based on its local information. The $\alpha$ and $\beta$ can be estimated to range in average from $0$ to $4$. They are strictly topology dependent, related to the network density and the number of simultaneously failed nodes and might change over time. A node could estimate their values real-time, or they can be decided in the design phase of the system. The $r$ value is strongly related with the timer used for the discovery. Its value depends on the network density, i.e. on the number of neighbors to discover. A node might retrieve the $r$ value during its preliminary phase operations. $r$ is added just once in (\ref{hbequation}) since we are assuming that the first node is going to bypass the failure thanks to the discovery and subsequent selection.\\
In summary, the adaptive failure recovery problem is solved by deriving the $X$ parameter that enables a node to make the adaptive selection between hop-based and multi-path recovery methods. (\ref{hblessmp}) shows the $X$ lower-bound when HB\hspace{3pt}$<$\hspace{3pt}MP.
\begin{equation}\label{hblessmp}
X > \dfrac{r + \alpha + 2 - \beta}{2}
\end{equation}
From (\ref{hblessmp}) it is clear that the larger the $r$ value is, the larger the $X$ is. This results in a movement of the failure node position closer to the head node location in order to achieve an overall HB recovery time more convenient compared to the MP one.
%

\section{Performance evaluation}\label{section4}
This section presents the BLE mesh network simulation model and its performance evaluation for compared methods.

\subsection{Simulation setup}
\begin{figure}[b]
	\vspace*{-4.5mm}
	\centerline{\includegraphics[trim=1.6cm 1.9cm 1.6cm 1.5cm,clip,width=6.2cm]{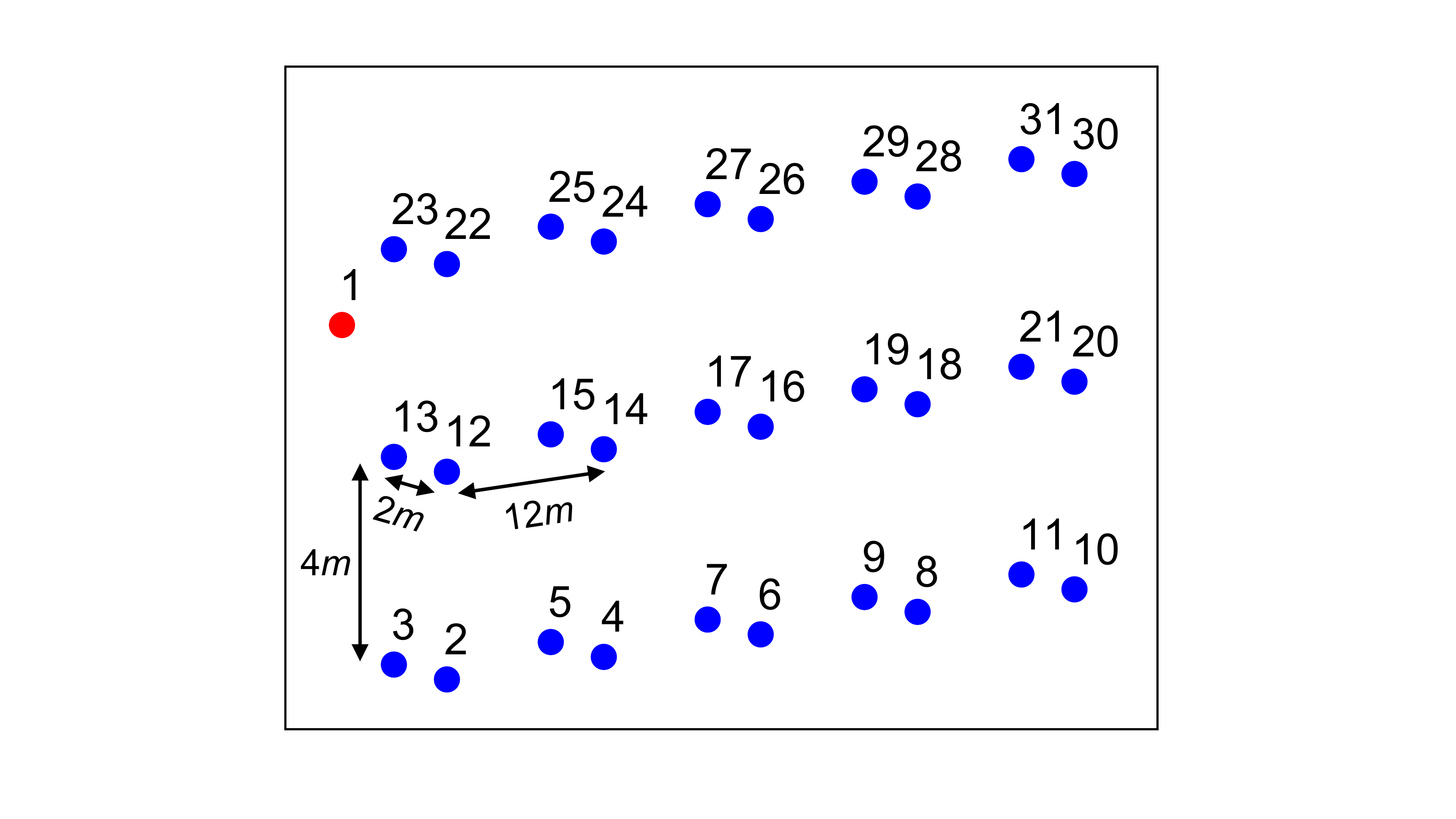}}
	\caption{Network topology in this study: \#1 head node; \#(2-31) BLE nodes.}
	\label{topologyexample}
\end{figure}

The network evaluated in this study is shown in Figure \ref{topologyexample}. It consists of a 3-floors topology which would simulate an indoor office or a student dormitory environment. Each floor is 4 $meters$ high and contains 10 nodes arranged in a 2x5 grid of 2x48 $meters$ with a distance of 12 $meters$ between two consecutive neighbors in each line and 1.2 $meters$ from the ground. One head node is located on the central floor ceiling.\\
The channel modeling exploits the BLE physical signal with Gaussian Frequency Shift Keying modulation. The International Telecommunication Union (ITU) indoor propagation model is used to define the path loss (PL) equation (\ref{itupathloss}) \cite{ituchannel}. It considers the transmission frequency ($f$), the distance between transmitter and receiver ($d$) with a power loss coefficient (\textit{N}) equal to 22, typical value for a commercial area in 2$GHz$ band, and a floor penetration loss ($P_f$) of 6 $dB$ per floor ($n$). The resulting path loss (PL) equation is (\ref{itupathloss}).
\begin{equation}\label{itupathloss}
PL_{d_0\rightarrow d} = 20log(f) + Nlog(d) + P_f(n) - 28
\end{equation}
A path loss threshold of 70 $dB$ determines if two nodes are considered as neighbors and can communicate directly and reliably. Additional parameters are summarized in Table \ref{simparameters}.\\
\begin{table}[t]	
	\caption{Simulation parameters.}
	\def\arraystretch{1.1}
	\begin{center}
		\begin{tabular}{|l|c|}
			\hline
			\textbf{Parameter} & \textbf{Value} \\ \hline
			$\tau_{AI}$ for head node & 100 $ms$ \\ \hline
			$\tau_{AI}$ for BLE nodes & 1 $second$ \\ \hline
			$\tau_{SW}$ equal to $\tau_{SI}$ & 10 $ms$ \\ \hline
			Discovery timer & 3 $seconds$ \\ \hline
			Reconnection timer & 6 $seconds$ \\ \hline
			Broadcast timer & 1.5 $seconds$ \\ \hline
			$\alpha$, $\beta$ (\ref{hblessmp}) & 0 \\ \hline
			$\gamma$ (\ref{hblessmp}) & 0.5 $seconds$ \\ \hline 
			$r$ (\ref{hblessmp}) & 2.3 $seconds$ \\ \hline
			Packet size & 240 $bits$ \\ \hline
		\end{tabular}
		\label{simparameters}
	\end{center}
	\vspace*{-4.5mm}
\end{table}
Each simulation run consists of the first phase of 1 $second$ long where all the nodes randomly wake up and start the advertisement cycle. After that, the nodes start performing the current simulation operations and the traffic is generated.\\
The first use case evaluates the preliminary phase performance, i.e. the neighbor discovery and the greedy search algorithm.\\
In the second and third use cases we compare the proposed disjoint multi-path method with a flooding routing technique with cache. In this flooding technique, the source of a message broadcasts the packet to all its 1-hop neighbors. Each neighbor re-broadcasts the message again if it is not the destination of the packet. The nodes maintain a cache in order not to deliver the same message twice. The system handles the broadcast of a packet in a sequentially unicast way creating the connections one at a time \cite{pat2}. In the proposed approach we assume that each node has performed the preliminary phase in advance and it has five disjoint paths stored in the routing tables.\\
The background traffic added in the second case consists of a fixed number of uplink packets equal to 1, 2 or 3 generated by different nodes at random in the network within a 1 $second$ window and they may or not all require an acknowledgment.\\
The third case tests a failure situation. The background traffic consists of one uplink packet generated by a node at random with a fixed hop-count of 5. A failed node is injected along the path that the message should follow triggering a failure recovery. The failed node hop-count varies between 1 and 4.\\
The main Key Performance Indicators (KPI) analyzed are latency and power consumption. The latency considers the average time needed to perform the end-to-end delivery of one data message from the original source till the final destination including any possible acknowledgment. The power consumption includes the average instantaneous current needed to perform the operations. The energy model considers the power consumption of a radio transceiver, which covers dominant energy cost in a wireless device. The values are based on the Texas Instrument Bluetooth chipset \textit{CC2642} \cite{chiptexas} at 3.3 $V$.

\subsection{Results analysis}
The simulation results for the first case are shown in Table \ref{firstcaseresults}. The neighbor discovery phase takes in average 5.5 $seconds$ which includes the timer of 3 $seconds$ plus around 2.5 $seconds$ for a node to retrieve data from 11 to 17 neighbors based on its location in the network. The overall phase, which includes discovery and paths creation, takes around 50 $seconds$ to create 5 disjoint paths through the GSA algorithm.\\
\begin{table}[t]	
	\caption{Preliminary phase results}
	\def\arraystretch{1.1}
	\begin{center}
		\begin{tabular}{|l|c|c|}
			\hline
			\textbf{Metric}       & \textbf{Discovery} & \textbf{All phase} \\ \hline
			Average delay per node ($s$)  &  5.52                       & 49.33                  \\ \hline
			Maximum delay per node ($s$)  &  9.19                       & 83.59                  \\ \hline
			Average node power consumption ($mW$)       & N/A                     & 8.05                    \\ \hline
		\end{tabular}
		\label{firstcaseresults}
	\end{center}
	\vspace*{-4.5mm}
\end{table}
\begin{figure}[b]
	\vspace*{-4.5mm}
	\centerline{\includegraphics[trim=4.2cm 6.9cm 3.8cm 6.9cm,clip,width=6.0cm]{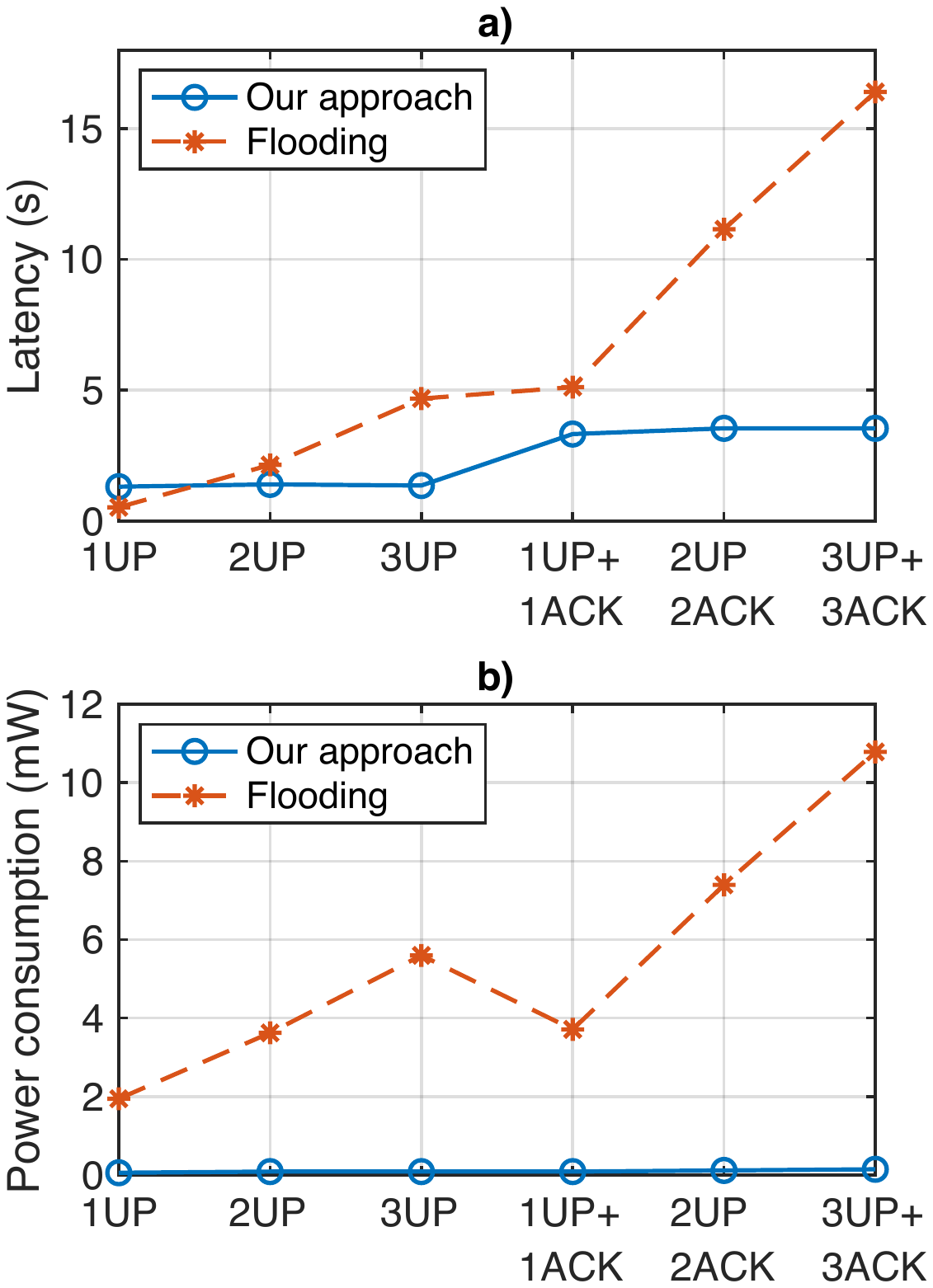}}
	\caption{Second case results of a normal situation for our approach and flooding: a) average latency for the end-to-end delivery of one uplink packet including its acknowledgment, if present; b) average power consumption in the first minute of simulation of one BLE node in the network.}
	\label{resultsnormal}
\end{figure}
\begin{figure}[t]
	\centerline{\includegraphics[trim=4.2cm 6.9cm 3.8cm 6.9cm,clip,width=6.0cm]{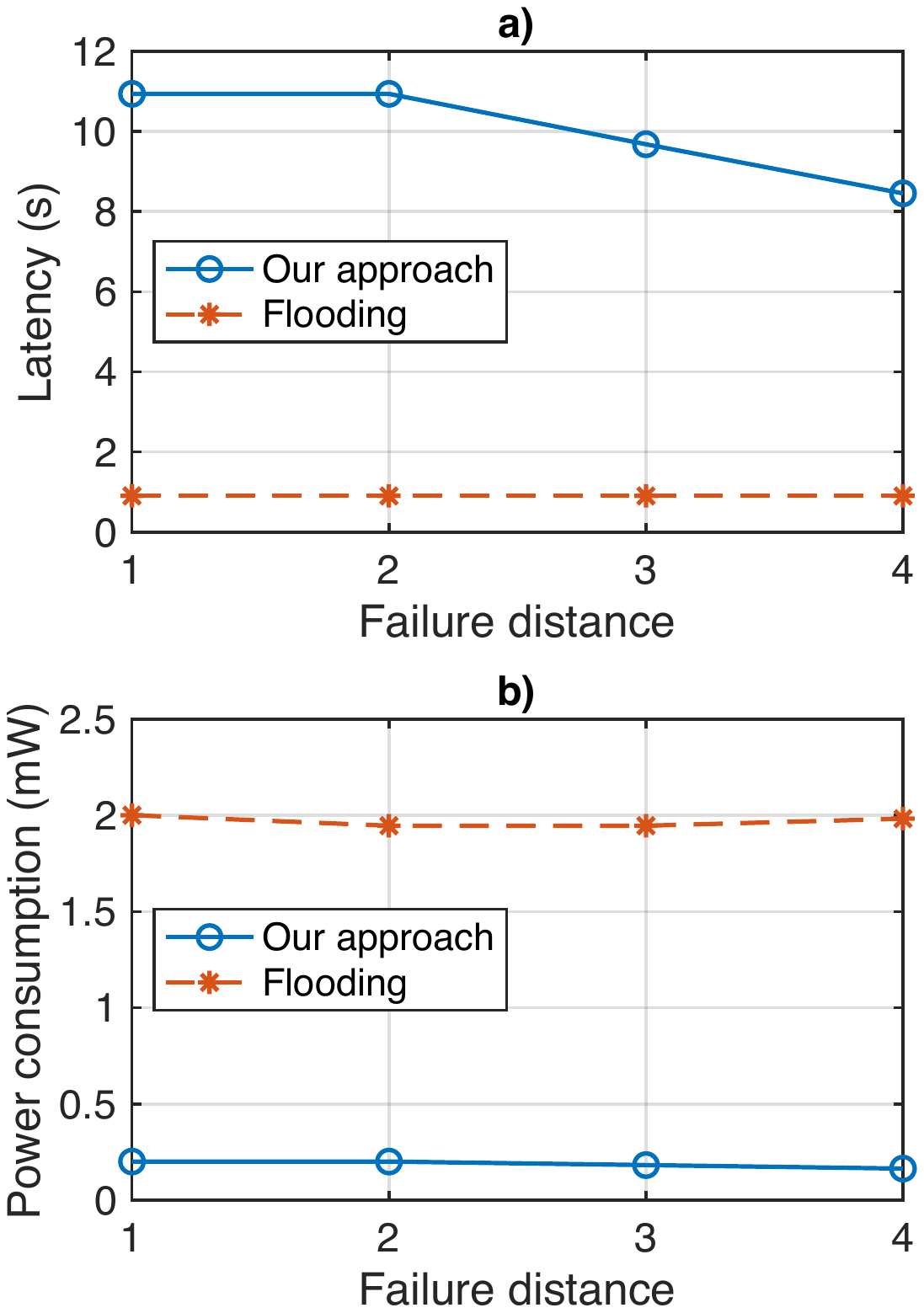}}
	\caption{Third case results of a failure situation for our approach and flooding: a) average latency for the end-to-end delivery of one packet; b) average power consumption in the first minute of simulation of one BLE node.}
	\label{resultsfailure}
	\vspace*{-4.5mm}
\end{figure}
Figure \ref{resultsnormal} presents the results for the second case. The overall latency increases when more packets are exchanged. In particular, this happens in the presence of acknowledgments since the directions of these packets are opposite, i.e. downlink, generating more congestion and queuing delay in the network. The power consumption follows the same trend as the latency. The overall performance are better in our proposed approach.\newline The flooding technique presents a higher queuing delay problem since the packets are served sequentially. The only exception is the first scenario with only one uplink packet since the message is promptly served and can reach the final destination from all directions without any queuing delay. However, the flooding power consumption is much higher in all scenarios due to the high volume of redundant messages exchanged.\\
The results for the third case are shown in Figure \ref{resultsfailure}. The latency in our proposed approach is much higher compared to the flooding one since the latter does not exploit any failure recovery mechanism, but the packets are always sent to all the nodes. However, the flooding power consumption is ten times more compared to our methodology because of all the redundant messages. We should notice that these are the results when a failure is discovered for the first time. In our approach a node will exploit the new selected route to deliver future data avoiding the failed spot and improving the performance.\\ 
\begin{figure}[b]
	\vspace*{-4.5mm}
	\centerline{\includegraphics[trim=2.0cm 8.9cm 1.8cm 9.4cm,clip,width=6.5cm]{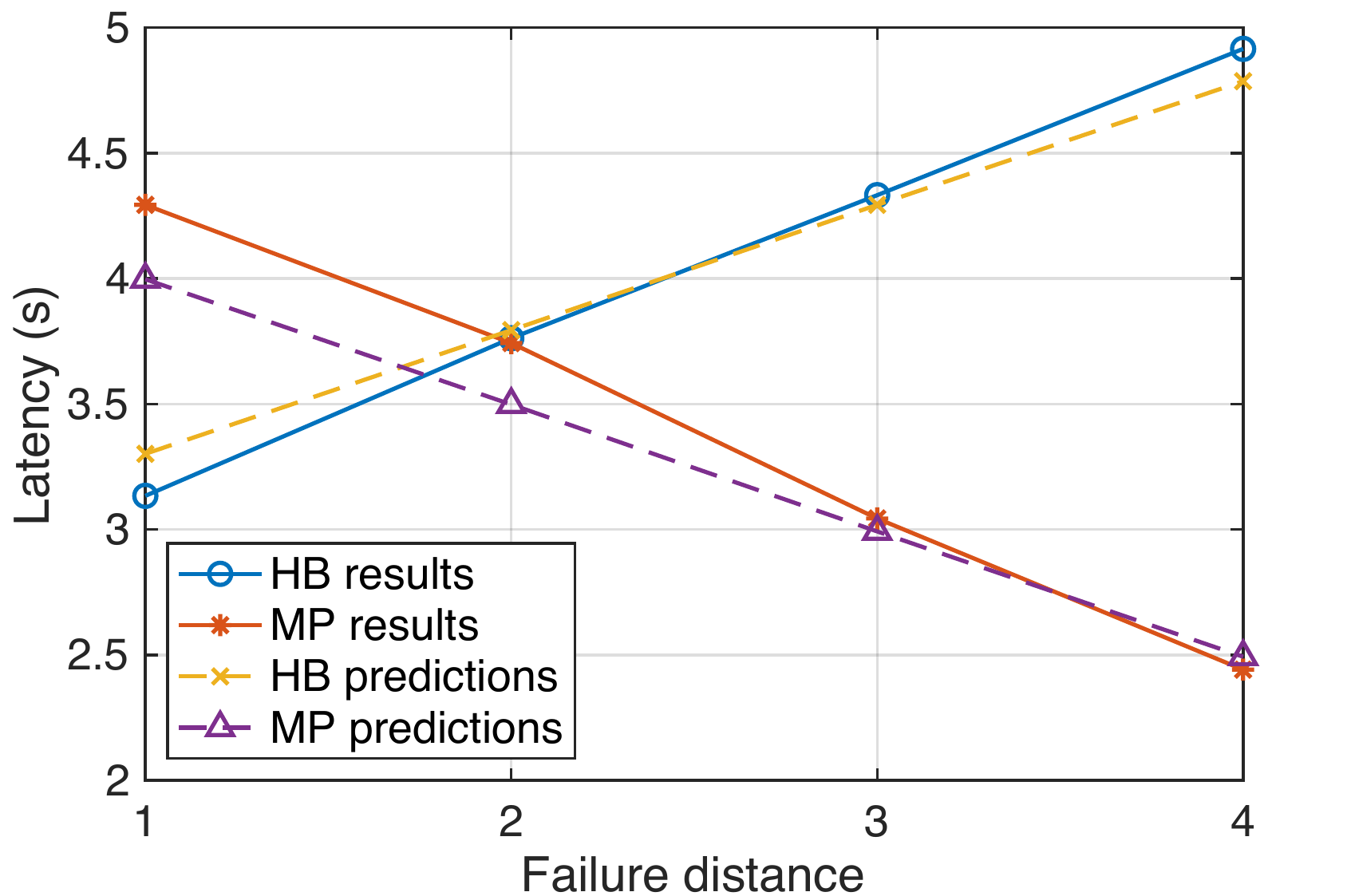}}
	\caption{Packet latency results and predictions of adaptive failure recovery for hop-distance (HB) and multi-path (MP) approaches with 4 failure distances.}
	\label{resultsadaptive}
\end{figure}
Additionally, Figure \ref{resultsadaptive} compares the failure recovery predictions (HB and MP) with the simulation results upon a failure detection. The estimated HB latency predictions increase with the increasing failure distance since $Z-X$ is higher.\linebreak On the contrary, the multi-path results improve since $X$ becomes smaller. The simulation results confirm the prediction model and find the switch point of the two recovery methods when a failure happens at 2-hops from the head in this network study. Beyond that point, the MP method becomes better.

\section{Conclusion}\label{section5}
In this paper we propose and demonstrate robust, reliable and low-power methodologies for wireless networks. These techniques include a preliminary phase and two failure recovery methods. The preliminary phase consists of a neighbor discovery to retrieve the 1-hop neighbors list and their hop-count distances towards the head, and a Greedy Search Algorithm to build a series of disjoint paths. The failure recoveries proposed are a multi-path approach, which exploits the existing routes, and a hop-distance based which relies on the neighbors' hop-counts to find an alternative path. The correctness, robustness and improvements of the proposed methods are evaluated with different test case in simulations and compared with a well-known flooding technique. Their cost difference and effectiveness are depending on the topology and failure scenario. The results show that the proposed methods improve the network performance in terms of packet latency and power consumption compared to a flooding technique and provide an adaptive and effective failure recovery in BLE mesh networks.


\nocite{*}
\bibliographystyle{ieeetr}
\bibliography{./BLE_Bibliography}

\end{document}